# Theory of Three-Pulse Photon Echo Spectroscopy with Dual Frequency Combs

JONGGU JEON,[1] JUNWOO KIM,[1] TAI HYUN YOON,[1,2] AND MINHAENG CHO[*,1,3]

[1]*Center for Molecular Spectroscopy and Dynamics, Institute for Basic Science (IBS), Seoul 02841, Republic of Korea*
[2]*Department of Physics, Korea University, Seoul 02841, Republic of Korea*
[3]*Department of Chemistry, Korea University, Seoul 02841, Republic of Korea*
[*]*E-mail: mcho@korea.ac.kr*

**Abstract:** A theoretical analysis is carried out for the recently developed three-pulse photon echo spectroscopy employing dual frequency combs (DFC) as the light sources. In this method, the molecular sample interacts with three pulse trains derived from the DFC and the generated third-order signal is displayed as a two-dimensional (2D) spectrum that depends on the waiting time introduced by employing asynchronous optical sampling method. Through the analysis of the heterodyne-detected signal interferogram using a local oscillator derived from one of the optical frequency combs, we show that the 2D spectrum closely matches the spectrum expected from a conventional approach with four pulses derived from a single femtosecond laser pulse and the waiting time between the second and third field-matter interactions is given by the down-converted detection time of the interferogram. The theoretical result is applied to a two-level model system with solvation effect described by solvatochromic spectral density. The model 2D spectrum reproduces spectral features such as the loss of frequency correlation, dephasing, and spectral shift as a function of the population time. We anticipate that the present theory will be the general framework for quantitative descriptions of DFC-based nonlinear optical spectroscopy.

## 1. Introduction

Three-pulse photon echo (3PPE) spectroscopy [1-4] is one of the most widely used two-dimensional (2D) spectroscopy techniques that can overcome many limitations of linear and time-resolved one-dimensional spectroscopy [5-8]. For example, 2D spectroscopy can resolve congested spectral features in 2D frequency space, distinguish homogeneous and inhomogeneous line-broadening mechanisms, and detect couplings between different optical transitions. In 3PPE spectroscopy, the molecular system interacts with three coherent laser pulses in a non-collinear four wave mixing scheme and the signal in a specific phase-matching direction is heterodyne-detected with a local oscillator field. The signal interferogram obtained from this measurement is then Fourier transformed over two time variables $\tau$ and $t$, representing the delays between pulses 1 and 2 and between pulse 3 and the detection time, respectively, to obtain the 2D spectrum in the conjugate frequency variables $\omega_\tau$ and $\omega_t$. The spectrum parametrically depends on the waiting or population time $T$ between pulses 2 and 3 and thereby conveys information on molecular dynamics and chemical and biological processes occurring over $T$.

Currently, the experimental feasibility of the 3PPE spectroscopy by using dual frequency combs (DFC) as the light sources is under investigation [9]. A frequency comb consists of a periodic and coherent pulse train with a specific phase between the optical carrier wave and the peak of the pulse envelope [10, 11]. By controlling the shift in this phase between successive pulses separated by $\Delta T = 2\pi/\omega_r$, which is called the carrier-envelope offset (CEO) phase $\Delta\phi_{ceo}$, it becomes possible to obtain equally spaced spectral peaks or combs with frequencies of $(\Delta\phi_{ceo}/2\pi + n)\omega_r$, i.e., integer-multiples of the repetition frequency $\omega_r$ plus an offset determined by the CEO phase. DFC spectroscopy [12-16] employs two frequency combs with a slight mismatch $\delta\omega_r = \omega_{r1} - \omega_{r2}$ in their repetition frequencies and this enables the asynchronous optical sampling (ASOPS) [14, 17], that is, an automatic scanning of delay times between two pulses arriving at the sample. This approach also enables the detection of optical resonances using radio frequency (RF) electronics through the down-conversion of the signal frequency from the optical to the RF range. DFC has also been applied to various nonlinear spectroscopy studies [16, 18-22]. Compared to typical nonlinear spectroscopy, DFC nonlinear spectroscopy provides fast data acquisition, high frequency resolution and relatively simple instrumentation based on the unique properties of DFC.

In our DFC 3PPE spectroscopy [9], four pulse trains derived from two frequency combs (combs 1 and 2) with slightly different repetition frequencies are used. Comb 1 is split into two and these two pulse trains arrive at the sample with a time delay $\tau_1$ and different propagation directions $\mathbf{k}_1$ and $\mathbf{k}_2$. The pulse train of comb 2 then arrives at the sample in yet another direction $\mathbf{k}_3$. Since the first two pulse trains are derived from comb 1, they have a fixed time gap $\tau_1$ which is chosen to be much smaller than the repetition period $\Delta T_1 = 2\pi/\omega_{r1}$ of comb 1. In contrast, the third pulse train comes from comb 2 and therefore is delayed by $\Delta T_2 - \Delta T_1 = 2\pi\delta\omega_r/(\omega_{r1}\omega_{r2})$ from the first two pulses after each repetition. Since the time between the second and third pulses is the population time, ASOPS is realized for the population time in this method [22]. The signal generated from the three field-matter interactions is heterodyne-detected with a local oscillator which is comb 2 delayed by $\tau_2$ and redirected in $\mathbf{k}_s = -\mathbf{k}_1 + \mathbf{k}_2 + \mathbf{k}_3$ direction. To obtain the 2D spectrum, the measurement is repeated with varying $\tau_1$ and $\tau_2$ up to the coherence decay time (~100 fs) and the signal interferogram is 2D Fourier transformed over $\tau_1$ and $\tau_2$. A typical frequency comb has the following parameters: the carrier frequency $\omega_{cj} \sim 1\,\text{PHz}$, $\omega_{rj} \sim 1\,\text{GHz}$, $\delta\omega_r \sim 1\,\text{kHz}$. (Throughout this paper, we express the numerical value of an angular frequency $\omega$ in either Hz or cm$^{-1}$ units that correspond to $\nu = \omega/(2\pi)$ and $\bar{\nu} = \omega/(2\pi c)$, respectively, where $c$ is the speed of light.) Therefore, the effective interval of the waiting time measurement is $2\pi\delta\omega_r/(\omega_{r1}\omega_{r2}) \sim 1\,\text{fs}$ while the laboratory time of measurement interval is $\Delta T_1 \sim \Delta T_2 \sim 1\,\text{ns}$. The ratio of these two quantities, $\delta\omega_r/\omega_{rj} \sim 10^{-6}$, is called the frequency down-conversion factor of DFC spectroscopy. This also alludes to the similar down-conversion relation $T_w \sim (\delta\omega_r/\omega_{rj})t \sim 10^{-6}t$ between the measurement time $t$ of the signal interferogram and the waiting time $T_w$.

In this paper, we refine the heuristic consideration presented above through a theoretical analysis of DFC 3PPE spectroscopy and aim to place the method on a solid theoretical basis. At the outset, the present approach is similar to our previous theory of DFC two-pulse photon echo spectroscopy [23] in fundamental ideas and extends it to take into account the presence of an additional pulse train and the extra time variable arising from it. Through the nonlinear response function formalism combined with a proper description of comb fields including the finite pulse width effect, we will show that the measurement time of the signal is indeed connected to the population time of 3PPE by the down-conversion factor. The theoretical 2D spectrum is calculated for a two-level model system (2LS) where the solvent effect is incorporated by introducing a model spectral density of the chromophore-bath coupling [8, 24].

In the next section, we present the theory of DFC 3PPE spectroscopy. We calculate and present the theoretical 2D spectrum for a 2LS in Section 3. Finally, a brief summary and concluding remarks are given in Section 4.

## 2. Theory

*A. Experimental Configuration*

We first specify the experimental condition of the DFC 3PPE spectroscopy for which our theory is developed [9]. We employ two frequency combs, each characterized by the repetition frequency $\omega_{rj}$ and the carrier frequency $\omega_{cj}$ ( $j=1,2$ ). For a coherent pulse train, they are related to each other by the following relation [10, 23, 25]

$$\omega_{cj} = n_{cj}\omega_{rj} + \omega_{ceo,j}, \tag{1}$$

where $n_{cj}$ is the integer mode number of the carrier frequency and $\omega_{ceo,j} = \Delta\phi_{ceo,j}\omega_{rj}/(2\pi)$ is the carrier-envelope offset (CEO) frequency defined in terms of the CEO phase shift $\Delta\phi_{ceo,j}$ between successive pulses of the $j$th pulse train. To enable ASOPS, we introduce a slight offset between the repetition frequencies of the two combs defined as

$$\delta\omega_r = \omega_{r1} - \omega_{r2}. \tag{2}$$

In this case, the two carrier frequencies do not coincide in general even if $\Delta\phi_{ceo,j}$ ( $j=1,2$ ) are precisely controlled to be identical [23] and their difference is denoted as

$$\delta\omega_c = \omega_{c1} - \omega_{c2}. \tag{3}$$

The pulse train of the $j$th comb is represented by the following electric field [23]

$$\mathcal{E}_j(\mathbf{r},t) = \frac{1}{2}\left[E_j(\mathbf{r},t) + \text{c.c.}\right],$$
$$E_j(\mathbf{r},t) = e^{i\mathbf{k}_j(\omega_{cj})\cdot\mathbf{r} - i\omega_{cj}t}\sum_{n=-\infty}^{\infty} A_{nj} e^{-in\omega_{rj}t}, \tag{4}$$

where c.c. denotes complex conjugate, $\mathbf{k}_j(\omega_{cj})$ is the wave vector of the $j$th comb, and $A_{nj}$ is the Fourier coefficient of the pulse envelope function $A_j(t)$ of the $j$th comb, which is given by

$$A_{nj} = \frac{\omega_{rj}}{2\pi}\int_{-\infty}^{\infty} dt\, A_j(t) e^{in\omega_{rj}t}. \tag{5}$$

The comb spectrum obtained by Fourier transform of Eq. (4) has peaks at the following frequencies

$$\omega = \omega_{cj} + n\omega_{rj} = \omega_{ceo,j} + m\omega_{rj} \quad (m,n = \text{integers}), \tag{6}$$

where the second equality comes from Eq. (1).

In the DFC 3PPE spectroscopy, the sample interacts with three sets of pulse trains: (i) comb 1 in $\mathbf{k}_1$ direction, (ii) comb 1 delayed by $\tau_1$ from the original comb 1 field and redirected in $\mathbf{k}_2$ direction, (iii) comb 2 in $\mathbf{k}_3$ direction. The waiting time $T_w$ between the second and third interactions is scanned automatically by the pulse trains of the combs 1 and 2 that have slightly different pulse repetition rates as given in Eq. (2). The signal field in the direction of $-\mathbf{k}_1 + \mathbf{k}_2 + \mathbf{k}_3$ is heterodyne-detected with a time-delayed comb 2 field as the local oscillator (LO), which is redirected in the signal direction and is delayed by $\tau_2$ from the third interaction. Note that $\tau_1$ and $\tau_2$ are explicitly controlled and scanned by, e.g., mechanical delay stages over a range $0 \leq \tau_1, \tau_2 \leq \tau_{\max}$ where $\tau_{\max}$ is of the order of coherence decay time and therefore is much shorter than the pulse repetition periods $\Delta T_j = 2\pi/\omega_{rj}$ of the two combs. In contrast, the waiting time $T_w$ is determined implicitly by the repetition frequencies of the two combs and their offset frequency $\delta\omega_r$ given in Eq. (2) as will be shown below. In practice, the time origin in the experiment is set to be the time when the second and third pulses maximally overlap.

With this proposition, the total electric field incident on the sample is the superposition of the three sets of pulse trains mentioned above

$$\mathcal{E}(\mathbf{r},t) = \mathcal{E}_1(\mathbf{r},t) + \mathcal{E}_2(\mathbf{r},t) + \mathcal{E}_3(\mathbf{r},t), \tag{7}$$

where $\mathcal{E}_1(\mathbf{r},t)$ and $\mathcal{E}_2(\mathbf{r},t)$ are derived from the comb 1 field and $\mathcal{E}_3(\mathbf{r},t)$ is the comb 2 field. They can be written in terms of complex electric fields as in the first member of Eq. (4). Hereafter, we take all electric fields as scalar quantities assuming a common direction of polarization. From Eq. (4), the three complex electric fields are given as

$$\begin{aligned} E_1(\mathbf{r},t) &= e^{i\mathbf{k}_1\cdot\mathbf{r}-i\omega_{c1}(t+\tau_1)} \sum_{n=-\infty}^{\infty} A_{n1} e^{-in\omega_{r1}(t+\tau_1)} \\ E_2(\mathbf{r},t) &= e^{i\mathbf{k}_2\cdot\mathbf{r}-i\omega_{c1}t} \sum_{n=-\infty}^{\infty} A_{n1} e^{-in\omega_{r1}t} \\ E_3(\mathbf{r},t) &= e^{i\mathbf{k}_3\cdot\mathbf{r}-i\omega_{c2}t} \sum_{n=-\infty}^{\infty} A_{n2} e^{-in\omega_{r2}t} \end{aligned} \tag{8}$$

Note that the time argument of $t+\tau_1$ is used in $E_1(\mathbf{r},t)$ because $E_1(\mathbf{r},t)$ precedes $E_2(\mathbf{r},t)$ by $\tau_1$. In addition, $E_2(\mathbf{r},t)$ and $E_3(\mathbf{r},t)$ overlap at $t=0$ (and, formally speaking, again at $t \simeq 2\pi n/\delta\omega_r$ for integer $n$, even though such a long time behavior is not needed here) as required by the above choice of time origin. The LO field used in the heterodyne detection of signal can be similarly written as follows

$$E_{\mathrm{LO}}(\mathbf{r},t) = e^{i\mathbf{k}_s\cdot\mathbf{r}-i\omega_{c2}(t-\tau_2)} \sum_{n=-\infty}^{\infty} A_{n2} e^{-in\omega_{r2}(t-\tau_2)}, \tag{9}$$

because it is assumed to be delayed by $\tau_2$ from the comb 2 field $E_3(\mathbf{r},t)$ and is redirected in the following direction to allow it to interfere with the 3PPE signal field,

$$\mathbf{k}_s = -\mathbf{k}_1 + \mathbf{k}_2 + \mathbf{k}_3. \tag{10}$$

## B. Third Order Polarization

The third order polarization generated by the incident electric field in Eq. (7) can be written in terms of the third order response function $S^{(3)}(t_3,t_2,t_1)$ as follows [26]

$$P^{(3)}(\mathbf{r},t) = \int_0^\infty dt_3 \int_0^\infty dt_2 \int_0^\infty dt_1 S^{(3)}(t_3,t_2,t_1) \mathcal{E}(\mathbf{r},t-t_3) \mathcal{E}(\mathbf{r},t-t_3-t_2) \mathcal{E}(\mathbf{r},t-t_3-t_2-t_1). \tag{11}$$

Since the signal is detected in the direction $\mathbf{k}_s = -\mathbf{k}_1 + \mathbf{k}_2 + \mathbf{k}_3$, the polarization component $P^{(3)}(\mathbf{k}_s,t)$ responsible for this signal has the following spatial dependence

$$P^{(3)}(\mathbf{k}_s,t) = P^{(3)}(t)\exp(i\mathbf{k}_s\cdot\mathbf{r}). \tag{12}$$

After expanding Eq. (11) using Eqs. (4), (7) and (8) and then selecting terms with the factor $\exp(i\mathbf{k}_s\cdot\mathbf{r})$, we can write Eq. (12) as follows

$$P^{(3)}(\mathbf{k}_s,t) = \frac{1}{8}\int_0^\infty dt_3 \int_0^\infty dt_2 \int_0^\infty dt_1 S^{(3)}(t_3,t_2,t_1)$$
$$\times \{E_1^*(\mathbf{r},t-t_3)E_2(\mathbf{r},t-t_3-t_2)E_3(\mathbf{r},t-t_3-t_2-t_1)$$
$$+ E_1^*(\mathbf{r},t-t_3)E_3(\mathbf{r},t-t_3-t_2)E_2(\mathbf{r},t-t_3-t_2-t_1)$$
$$+ E_2(\mathbf{r},t-t_3)E_1^*(\mathbf{r},t-t_3-t_2)E_3(\mathbf{r},t-t_3-t_2-t_1) \quad . \quad (13)$$
$$+ E_2(\mathbf{r},t-t_3)E_3(\mathbf{r},t-t_3-t_2)E_1^*(\mathbf{r},t-t_3-t_2-t_1)$$
$$+ E_3(\mathbf{r},t-t_3)E_1^*(\mathbf{r},t-t_3-t_2)E_2(\mathbf{r},t-t_3-t_2-t_1)$$
$$+ E_3(\mathbf{r},t-t_3)E_2(\mathbf{r},t-t_3-t_2)E_1^*(\mathbf{r},t-t_3-t_2-t_1)\}$$

According to Eq. (8), within a time span $0 \leq t < \pi/\delta\omega_r$, $E_1$ always arrives first at the sample followed by $E_2$, then by $E_3$. For $\pi/\delta\omega_r \leq t < 2\pi/\delta\omega_r$, this time ordering cannot be regarded as valid. However, we can simply disregard $t$ larger than the lifetime of the relevant excited state that is in general much shorter than $\pi/\delta\omega_r$. Nonetheless, the interaction times represented by the time arguments in each $E_k$ in Eq. (13) do not necessarily follow this order. For example, the first term in the integrand corresponds to the situation where $E_3$ interacts first with the sample at time $t-t_3-t_2-t_1$, $E_2$ interacts next at $t-t_3-t_2$, then $E_1$ at $t-t_3$, disregarding minor exceptions due to finite $\tau_1$ or $\tau_2$. From this consideration, only the sixth term above is consistent with the time ordering of the pulses and the remaining five terms will only contribute for very narrow time windows in $t_n$ ($n = 1, 2,$ and 3) set by the pulse widths, that is, when two or more pulses overlap significantly. Effectively, each of these five terms becomes non-negligible only if one or more of $\tau_1$, $\tau_2$, or the implicit waiting time $T_w$ is smaller than the pulse width and becomes a delta-function-like single point contribution in the impulsive limit. In particular, the fourth term, where only the interaction times of $E_2$ and $E_3$ are not consistent with the pulse time ordering, would be non-negligible only at small $T_w$ and therefore can be classified as coherent artifact [16, 22, 27]. In general, for a waiting time longer than the pulse width, this coherent artifact would contribute negligibly. In the following, we develop the theory including all sixth terms for completeness but, in the model calculation, take into account only the sixth term which is expected to predominantly contribute to the observed signal for waiting times longer than the incident pulse duration time.

With this precaution, we expand the integrand of Eq. (13) to obtain

$$P^{(3)}(\mathbf{k}_s,t) = \frac{1}{8}e^{i\mathbf{k}_s \cdot \mathbf{r}} \sum_{\alpha=1}^{6} \sum_{q,m,n=-\infty}^{\infty} c_{qmn}^{[\alpha]} e^{i\omega_{t,qmn}^{[\alpha]}t} e^{i\omega_{\tau 1,qmn}^{[\alpha]}\tau_1}$$
$$\times \int_0^\infty dt_3 \int_0^\infty dt_2 \int_0^\infty dt_1 S^{(3)}(t_3,t_2,t_1)e^{i(\omega_3^{[\alpha]}t_3+\omega_2^{[\alpha]}t_2+\omega_1^{[\alpha]}t_1)} \quad , \quad (14)$$
$$= \frac{1}{8}e^{i\mathbf{k}_s \cdot \mathbf{r}} \sum_{\alpha=1}^{6} \sum_{q,m,n=-\infty}^{\infty} c_{qmn}^{[\alpha]} \tilde{S}^{(3)}(\omega_{3,qmn}^{[\alpha]},\omega_{2,qmn}^{[\alpha]},\omega_{1,qmn}^{[\alpha]})e^{i\omega_{t,qmn}^{[\alpha]}t} e^{i\omega_{\tau 1,qmn}^{[\alpha]}\tau_1}$$

where we have introduced the frequency-domain response function or nonlinear susceptibility as

$$\tilde{S}^{(3)}(\omega_3,\omega_2,\omega_1) \equiv \int_0^\infty dt_3 \int_0^\infty dt_2 \int_0^\infty dt_1 S^{(3)}(t_3,t_2,t_1)e^{i(\omega_3 t_3+\omega_2 t_2+\omega_1 t_1)} \quad (15)$$

and the following notations

$$\omega_j^n = \omega_{cj} + n\omega_{rj}, \tag{16}$$

$$\begin{aligned}
c_{qmn}^{[1]} &= A_{q1}^* A_{m1} A_{n2}, & c_{qmn}^{[2]} &= A_{q1}^* A_{n1} A_{m2}, & c_{qmn}^{[3]} &= A_{m1}^* A_{q1} A_{n2}, \\
c_{qmn}^{[4]} &= A_{n1}^* A_{q1} A_{m2}, & c_{qmn}^{[5]} &= A_{m1}^* A_{n1} A_{q2}, & c_{qmn}^{[6]} &= A_{n1}^* A_{m1} A_{q2},
\end{aligned} \tag{17}$$

$$\begin{aligned}
\omega_{t,qmn}^{[1]} &= \omega_1^q - \omega_1^m - \omega_2^n, & \omega_{t,qmn}^{[2]} &= \omega_1^q - \omega_1^n - \omega_2^m, & \omega_{t,qmn}^{[3]} &= \omega_1^m - \omega_1^q - \omega_2^n, \\
\omega_{t,qmn}^{[4]} &= \omega_1^n - \omega_1^q - \omega_2^m, & \omega_{t,qmn}^{[5]} &= \omega_1^m - \omega_1^n - \omega_2^q, & \omega_{t,qmn}^{[6]} &= \omega_1^n - \omega_1^m - \omega_2^q,
\end{aligned} \tag{18}$$

$$\omega_{\tau 1,qmn}^{[1]} = \omega_{\tau 1,qmn}^{[2]} = \omega_1^q, \quad \omega_{\tau 1,qmn}^{[3]} = \omega_{\tau 1,qmn}^{[5]} = \omega_1^m, \quad \omega_{\tau 1,qmn}^{[4]} = \omega_{\tau 1,qmn}^{[6]} = \omega_1^n, \tag{19}$$

$$\begin{aligned}
\omega_{1,qmn}^{[1]} &= \omega_{1,qmn}^{[3]} = \omega_2^n, \quad \omega_{1,qmn}^{[2]} = \omega_{1,qmn}^{[5]} = -\omega_{1,qmn}^{[4]} = -\omega_{1,qmn}^{[6]} = \omega_1^n, \\
\omega_{2,qmn}^{[1]} &= \omega_1^m + \omega_2^n, \quad \omega_{2,qmn}^{[2]} = \omega_1^n + \omega_2^m, \quad \omega_{2,qmn}^{[3]} = -\omega_1^m + \omega_2^n, \\
\omega_{2,qmn}^{[4]} &= -\omega_1^n + \omega_2^m, \quad \omega_{2,qmn}^{[5]} = -\omega_{2,qmn}^{[6]} = -\omega_1^m + \omega_1^n, \\
\omega_{3,qmn}^{[\alpha]} &= -\omega_{t,qmn}^{[\alpha]} \quad (\alpha = 1,\cdots,6).
\end{aligned} \tag{20}$$

The electric field $E^{(3)}(\mathbf{k}_s,t)$ generated by $P^{(3)}(\mathbf{k}_s,t)$ is, within the slowly-varying-amplitude approximation, given as [26, 28]

$$E^{(3)}(\mathbf{k}_s,t) \propto i P^{(3)}(\mathbf{k}_s,t). \tag{21}$$

*C. Heterodyne-Detected Signal*

When the signal is heterodyne-detected with the LO which is the redirected and delayed comb 2 field, the signal field at the square detector is given as the superposition of $E_{\mathrm{LO}}$ and the third-order signal field $E^{(3)}$ as follows

$$I(t) = \left|E_{\mathrm{LO}}(\mathbf{k}_s,t) + E^{(3)}(\mathbf{k}_s,t)\right|^2 = \left|E_{\mathrm{LO}}(\mathbf{k}_s,t)\right|^2 + \left|E^{(3)}(\mathbf{k}_s,t)\right|^2 + 2\mathrm{Re}\left[E_{\mathrm{LO}}^*(\mathbf{k}_s,t)E^{(3)}(\mathbf{k}_s,t)\right]. \tag{22}$$

In general, $E^{(3)}$ is much weaker than $E_{\mathrm{LO}}$, making the second term on the right-hand side of this equation negligible. After subtracting the first term which is a known quantity, we obtain the last term which is a time-dependent interferogram containing information on the material response. Using Eqs. (9), (14), and (21), it can be written as

$$\begin{aligned}
2\mathrm{Re}&\left[E_{\mathrm{LO}}^*(\mathbf{k}_s,t)E^{(3)}(\mathbf{k}_s,t)\right] \propto 2\mathrm{Im}[E_{\mathrm{LO}}^*(\mathbf{k}_s,t)P^{(3)}(\mathbf{k}_s,t)] \\
&= \frac{1}{4}\mathrm{Im}\left[\sum_{\alpha=1}^{6}\sum_{p,q,m,n=-\infty}^{\infty} A_{p2}^* c_{qmn}^{[\alpha]} \tilde{S}^{(3)}(\omega_{3,qmn}^{[\alpha]},\omega_{2,qmn}^{[\alpha]},\omega_{1,qmn}^{[\alpha]}) e^{i\left[\omega_2^p + \omega_{t,qmn}^{[\alpha]}\right]t} e^{i\omega_{\tau 1,qmn}^{[\alpha]}\tau_1} e^{-i\omega_2^p \tau_2}\right].
\end{aligned} \tag{23}$$

The dependence of this interferogram on $t$ is determined by the frequency factors $\omega_2^p + \omega_{t,qmn}^{[\alpha]}$ in the exponent which can be written as follows using Eqs. (2), (16), and (18)

$$\begin{aligned}
\omega_2^p + \omega_t^{[1]} &= (q-m)\delta\omega_r + (p+q-m-n)\omega_{r2} \\
\omega_2^p + \omega_t^{[2]} &= (q-n)\delta\omega_r + (p+q-m-n)\omega_{r2} \\
\omega_2^p + \omega_t^{[3]} &= -(q-m)\delta\omega_r + (p-q+m-n)\omega_{r2} \\
\omega_2^p + \omega_t^{[4]} &= -(q-n)\delta\omega_r + (p-q-m+n)\omega_{r2} \\
\omega_2^p + \omega_t^{[5]} &= (m-n)\delta\omega_r + (p-q+m-n)\omega_{r2} \\
\omega_2^p + \omega_t^{[6]} &= -(m-n)\delta\omega_r + (p-q-m+n)\omega_{r2}
\end{aligned} \tag{24}$$

In the experiment, only the slowly oscillating interference terms are selectively detected with a low-pass filter or a slow-response detector. Under this condition, the terms with non-zero coefficients of $\omega_{r2}$ would exhibit high-frequency oscillation due to the relation $\omega_{r2} \approx 10^6 \, \delta\omega_r$, and therefore are not detected [23]. Then, only the terms with the following values of $p$ and associated frequency factors survive when detecting the signal in time $t$

$$
\begin{aligned}
\alpha = 1: \quad & p = -q + m + n, \quad \omega_2^p + \omega_t^{[1]} = (q-m)\delta\omega_r \\
\alpha = 2: \quad & p = -q + m + n, \quad \omega_2^p + \omega_t^{[2]} = (q-n)\delta\omega_r \\
\alpha = 3: \quad & p = q - m + n, \quad \omega_2^p + \omega_t^{[3]} = -(q-m)\delta\omega_r \\
\alpha = 4: \quad & p = q + m - n, \quad \omega_2^p + \omega_t^{[4]} = -(q-n)\delta\omega_r \\
\alpha = 5: \quad & p = q - m + n, \quad \omega_2^p + \omega_t^{[5]} = (m-n)\delta\omega_r \\
\alpha = 6: \quad & p = q + m - n, \quad \omega_2^p + \omega_t^{[6]} = -(m-n)\delta\omega_r
\end{aligned}
\tag{25}
$$

The heterodyne-detected signal in Eq. (23) can then be written as

$$
\begin{aligned}
2\,\mathrm{Re}\!\left[ E_{\mathrm{LO}}^*(\mathbf{k}_s,t) E^{(3)}(\mathbf{k}_s,t) \right] &\propto 2\,\mathrm{Im}[E_{\mathrm{LO}}^*(\mathbf{k}_s,t) P^{(3)}(\mathbf{k}_s,t)] \\
&= \frac{1}{4}\mathrm{Im}\!\left[ \sum_{L,M,N=-\infty}^{\infty} B_{LMN} e^{iL\delta\omega_r t} e^{i\omega_1^M \tau_1} e^{-i\omega_2^N \tau_2} \right]
\end{aligned}
\tag{26}
$$

where the coefficient $B_{LMN}$ is given by

$$
\begin{aligned}
B_{LMN} = A_{M,1}^* A_{M-L,1} A_{N,2}^* A_{L+N,2} \Big[ & \tilde{S}^{(3)}(-L\omega_{r1}+\omega_2^{L+N}, \omega_1^{M-L}+\omega_2^{L+N}, \omega_2^{L+N}) \\
& + \tilde{S}^{(3)}(-L\omega_{r1}+\omega_2^{L+N}, \omega_1^{M-L}+\omega_2^{L+N}, \omega_1^{M-L}) + \tilde{S}^{(3)}(-L\omega_{r1}+\omega_2^{L+N}, -\omega_1^M+\omega_2^{L+N}, \omega_2^{L+N}) \\
& + \tilde{S}^{(3)}(-L\omega_{r1}+\omega_2^{L+N}, -\omega_1^M+\omega_2^{L+N}, -\omega_1^M) + \tilde{S}^{(3)}(-L\omega_{r1}+\omega_2^{L+N}, -L\omega_{r1}, \omega_1^{M-L}) \\
& + \tilde{S}^{(3)}(-L\omega_{r1}+\omega_2^{L+N}, -L\omega_{r1}, -\omega_1^M) \Big]
\end{aligned}
\tag{27}
$$

### D. Two-Dimensional Spectrum

To obtain a 2D spectrum from the heterodyne-detected signal in Eq. (26), we first define a complex function $S(t;\tau_1,\tau_2)$ as

$$
S(t;\tau_1,\tau_2) = \frac{1}{8} \sum_{L,M,N=-\infty}^{\infty} B_{LMN} e^{iL\delta\omega_r t} e^{i\omega_1^M \tau_1} e^{-i\omega_2^N \tau_2} .
\tag{28}
$$

The imaginary part of $S(t;\tau_1,\tau_2)$, which we denote as $S_I(t;\tau_1,\tau_2)$, is directly related to the signal as follows

$$
2\,\mathrm{Im}[E_{\mathrm{LO}}^*(\mathbf{k}_s,t) P^{(3)}(\mathbf{k}_s,t)] = 2 S_I(t;\tau_1,\tau_2) .
\tag{29}
$$

We then perform the 2D Fourier transform of $S(t;\tau_1,\tau_2)$ with respect to $\tau_1$ and $\tau_2$ to obtain

$$
\begin{aligned}
\tilde{S}(t;\omega_{\tau 1},\omega_{\tau 2}) &\equiv \int_{-\infty}^{\infty} d\tau_1 \int_{-\infty}^{\infty} d\tau_2\, S(t;\tau_1,\tau_2) e^{i\omega_{\tau 1}\tau_1} e^{i\omega_{\tau 2}\tau_2} \\
&= \frac{1}{8} \sum_{L,M,N=-\infty}^{\infty} B_{LMN} e^{iL\delta\omega_r t} \int_{-\infty}^{\infty} d\tau_1 e^{i(\omega_{\tau 1}+\omega_1^M)\tau_1} \int_{-\infty}^{\infty} d\tau_2 e^{i(\omega_{\tau 2}-\omega_2^N)\tau_2} \\
&= \frac{\pi^2}{2} \sum_{L,M,N=-\infty}^{\infty} B_{LMN} e^{iL\delta\omega_r t} \delta(\omega_{\tau 1}+\omega_1^M) \delta(\omega_{\tau 2}-\omega_2^N).
\end{aligned}
\tag{30}
$$

From Eq. (29), the 2D signal can be defined as the 2D Fourier transform of $2S_I(t;\tau_1,\tau_2)$. It can be written in terms of $\tilde{S}(t;\omega_{\tau 1},\omega_{\tau 2})$ above as follows

$$2\tilde{S}_I(t;\omega_{\tau 1},\omega_{\tau 2}) \equiv 2\int_{-\infty}^{\infty}d\tau_1\int_{-\infty}^{\infty}d\tau_2\, S_I(t;\tau_1,\tau_2)e^{i\omega_{\tau 1}\tau_1}e^{i\omega_{\tau 2}\tau_2}$$
$$= \text{Im}\left[\tilde{S}(t;\omega_{\tau 1},\omega_{\tau 2}) + \tilde{S}(t;-\omega_{\tau 1},-\omega_{\tau 2})\right] - i\,\text{Re}\left[\tilde{S}(t;\omega_{\tau 1},\omega_{\tau 2}) - \tilde{S}(t;-\omega_{\tau 1},-\omega_{\tau 2})\right]. \quad (31)$$

From Eqs. (30) and (31), we find that the 2D signal $2\tilde{S}_I(t;\omega_{\tau 1},\omega_{\tau 2})$ has peaks at $(\omega_{\tau 1},\omega_{\tau 2}) = \pm(-\omega_1^M,\omega_2^N)$ for integers $M$ and $N$, displaying a 2D comb-like peak arrangement with comb spacings of $\omega_{r1}$ and $\omega_{r2}$ along the $\omega_{\tau 1}$ and $\omega_{\tau 2}$ axes, respectively. Moreover, the amplitude of the signal at a given frequency point $(\omega_{\tau 1},\omega_{\tau 2})$ is determined by $B_{LMN}$ with the following values of $M$ and $N$:

$$(\bar{M},\bar{N}) = \left((\mp\omega_{\tau 1} - \omega_{c1})/\omega_{r1},(\pm\omega_{\tau 2} - \omega_{c2})/\omega_{r2}\right), \quad (32)$$

where the upper signs come from $\tilde{S}(t,\omega_{\tau 1},\omega_{\tau 2})$ and the lower signs come from $\tilde{S}(t,-\omega_{\tau 1},-\omega_{\tau 2})$ in Eq. (31).

The 2D signal in Eq. (31) depends on the detection time variable $t$ according to Eq. (30). Because the $t$-dependence is present only in the factor $\exp(iL\delta\omega_r t)$, we focus on the summation over $L$ and approximate it as an integral over a frequency variable $\omega$ using the correspondence $\omega \leftrightarrow L\delta\omega_r$:

$$\sum_{L=-\infty}^{\infty} B_{LMN}e^{iL\delta\omega_r t} \simeq \frac{1}{\delta\omega_r}\int_{-\infty}^{\infty}d\omega\,\bar{B}_{MN}(\omega)e^{i\omega t}$$
$$= \frac{2\pi}{\delta\omega_r}\frac{1}{2\pi}\int_{-\infty}^{\infty}d\omega\,\bar{B}_{MN}(\omega)e^{-i\omega(-t)}, \quad (33)$$
$$= \frac{2\pi}{\delta\omega_r}\bar{\bar{B}}_{MN}(-t)$$

where $\bar{B}_{MN}(\omega) = B_{(\omega/\delta\omega_r)MN}$ and $\bar{\bar{B}}_{MN}(t)$ denotes the inverse Fourier transform of $\bar{B}_{MN}(\omega)$. Then, $\tilde{S}(t;\omega_{\tau 1},\omega_{\tau 2})$ in Eq. (30) can be rewritten as

$$\tilde{S}(t;\omega_{\tau 1},\omega_{\tau 2}) = \frac{\pi^3}{\delta\omega_r}\sum_{M,N=-\infty}^{\infty}\bar{\bar{B}}_{MN}(-t)\delta(\omega_{\tau 1}+\omega_1^M)\delta(\omega_{\tau 2}-\omega_2^N). \quad (34)$$

Similarly, the time-domain function $S(t;\tau_1,\tau_2)$ in Eq. (28) can be expressed as

$$S(t;\tau_1,\tau_2) = \frac{\pi}{4\delta\omega_r}\sum_{M,N=-\infty}^{\infty}\bar{\bar{B}}_{MN}(-t)e^{i\omega_1^M\tau_1}e^{-i\omega_2^N\tau_2} \quad (35)$$

As in our previous study [23], we also introduce the amplitude function $F(t;\omega_{\tau 1},\omega_{\tau 2})$ of the 2D signal as

$$F(t;\omega_{\tau 1},\omega_{\tau 2}) = \frac{\pi^3}{\delta\omega_r}\bar{\bar{B}}_{\bar{M}(\omega_{\tau 1})\bar{N}(\omega_{\tau 2})}(-t) \quad (36)$$

where $\bar{M}(\omega_{\tau 1})$ and $\bar{N}(\omega_{\tau 2})$ depend on $\omega_{\tau 1}$ and $\omega_{\tau 2}$, respectively, through the following relations

$$\begin{aligned}\bar{M}(\omega_{\tau 1}) &= -(\omega_{\tau 1}+\omega_{c1})/\omega_{r1}, \\ \bar{N}(\omega_{\tau 2}) &= (\omega_{\tau 2}-\omega_{c2})/\omega_{r2}\end{aligned} \quad (37)$$

which correspond to the upper set of signs in Eq. (32). Finally, the 2D spectrum $S_{2D}(t;\omega_{\tau 1},\omega_{\tau 2})$, which is the amplitude of the signal at a given point $(\omega_{\tau 1},\omega_{\tau 2})$ in the 2D frequency space, can be constructed using $F(t;\omega_{\tau 1},\omega_{\tau 2})$ as follows

$$S_{2D}(t;\omega_{\tau 1},\omega_{\tau 2}) = \text{Im}\left[F(t;\omega_{\tau 1},\omega_{\tau 2})+F(t;-\omega_{\tau 1},-\omega_{\tau 2})\right]-i\,\text{Re}\left[F(t;\omega_{\tau 1},\omega_{\tau 2})-F(t;-\omega_{\tau 1},-\omega_{\tau 2})\right] \quad (38)$$

in analogy with Eq. (31). Note that $F(t;-\omega_{\tau 1},-\omega_{\tau 2})$ in this equation is given by $\left(\pi^3/\delta\omega_r\right)\bar{\bar{B}}_{\bar{M}(-\omega_{\tau 1})\bar{N}(-\omega_{\tau 2})}(-t)$ according to Eq. (36).

*E. Waiting Time Dependence of 2D Spectrum*

We now further simplify the 2D spectrum derived above and investigate its dependence on the waiting time $T_w$. We first consider $\bar{B}_{MN}(\omega)$ introduced above in Eq. (33) and write it explicitly as follows using Eq. (27) and the correspondence $\omega \leftrightarrow L\delta\omega_r$

$$\begin{aligned}\bar{B}_{MN}(\omega) = G_{MN}(\omega)\big[&\tilde{S}^{(3)}(\omega_2^N-\omega,\omega_1^M+\omega_2^N-\omega,\omega_2^N+\omega_{r2}\omega/\delta\omega_r) \\ +&\tilde{S}^{(3)}(\omega_2^N-\omega,\omega_1^M+\omega_2^N-\omega,\omega_1^M-\omega_{r1}\omega/\delta\omega_r) \\ +&\tilde{S}^{(3)}(\omega_2^N-\omega,-\omega_1^M+\omega_2^N+\omega_{r2}\omega/\delta\omega_r,\omega_2^N+\omega_{r2}\omega/\delta\omega_r) \\ +&\tilde{S}^{(3)}(\omega_2^N-\omega,-\omega_1^M+\omega_2^N+\omega_{r2}\omega/\delta\omega_r,-\omega_1^M) \\ +&\tilde{S}^{(3)}(\omega_2^N-\omega,-\omega_{r1}\omega/\delta\omega_r,\omega_1^M-\omega_{r1}\omega/\delta\omega_r) \\ +&\tilde{S}^{(3)}(\omega_2^N-\omega,-\omega_{r1}\omega/\delta\omega_r,-\omega_1^M)\big]\end{aligned} \quad (39)$$

where we have introduced the function $G_{MN}(\omega)$ as

$$G_{MN}(\omega) = A^*_{M,1} A_{M-\omega/\delta\omega_r,1} A^*_{N,2} A_{\omega/\delta\omega_r+N,2}. \quad (40)$$

For a pulse envelope with finite width, the Fourier coefficient $A_{nj}$ of the envelope function is nonzero only for a limited range of $n$ centered at zero. This effectively determines the range of $\omega$ to consider in Eq. (39). To be concrete, we assume that the pulse envelope is given by a Gaussian function of the form

$$A_j(t) = \frac{1}{\sigma_j\sqrt{2\pi}} e^{-t^2/(2\sigma_j^2)}. \quad (41)$$

Then, $A_{nj}$ can be written as follows using Eq. (5)

$$A_{nj} = \frac{\omega_{rj}}{2\pi} e^{-n^2\sigma_j^2\omega_{rj}^2/2} \quad (42)$$

and $G_{MN}(\omega)$ in Eq. (40) becomes

$$G_{MN}(\omega) = \frac{\omega_{r1}^2 \omega_{r2}^2}{16\pi^4} e^{-\sigma_1^2 \omega_{r1}^2 (M - \omega/(2\delta\omega_r))^2 - \sigma_2^2 \omega_{r2}^2 (N + \omega/(2\delta\omega_r))^2 - (\omega^2/(2\delta\omega_r)^2)(\sigma_1^2 \omega_{r1}^2 + \sigma_2^2 \omega_{r2}^2)}. \tag{43}$$

We define the range of $x$ in which the function $G(x) = \exp(-\sigma_j^2 \omega_{rj}^2 x^2/2)$ is non-negligible as

$$-a/(\sigma_j \omega_{rj}) \le x \le a/(\sigma_j \omega_{rj}) \quad (a > 0) \tag{44}$$

using the parameter $a$ (~ 2 to 5) that controls the range and further note that $\sigma_1 \simeq \sigma_2$ and $\omega_{r1} \simeq \omega_{r2}$. Then, the corresponding ranges of $\omega/\delta\omega_r$, $M$, and $N$, over which $G_{MN}(\omega)$ in Eq. (43) is non-negligible, can be found as

$$\begin{aligned}-a/(\sigma_j \omega_{rj}) \le \omega/\delta\omega_r \le a/(\sigma_j \omega_{rj}), \\ -3a/(\sqrt{2}\sigma_j \omega_{rj}) \le M, N \le 3a/(\sqrt{2}\sigma_j \omega_{rj}).\end{aligned} \tag{45}$$

Therefore, the range of $\omega$ where $\bar{B}_{MN}(\omega)$ is non-negligible is given by

$$-a\delta\omega_r/(\sigma_j \omega_{rj}) \le \omega \le a\delta\omega_r/(\sigma_j \omega_{rj}). \tag{46}$$

Typical magnitudes of the repetition frequency variables are $\omega_{rj} \sim 1$ GHz and $\delta\omega_r/\omega_{rj} \sim 10^{-6}$. Then, choosing $a \sim 3$ and $\sigma_j = 10$ fs, we have $a/(\sigma_j \omega_{rj}) \sim 10^5$ and $a\delta\omega_r/(\sigma_j \omega_{rj}) \sim 0.3$ GHz or 0.01 cm$^{-1}$. Therefore, the extremal values of $\omega$ to consider are somewhat smaller than $\omega_{rj}$. Based on this analysis, we can safely simplify the arguments in the response functions as follows:

$$\begin{aligned}\omega_2^N - \omega &\simeq \omega_2^N, \\ \omega_1^M + \omega_2^N - \omega &\simeq \omega_1^M + \omega_2^N.\end{aligned} \tag{47}$$

Then, to a very good approximation, $\bar{B}_{MN}(\omega)$ in Eq. (39) can be written as

$$\begin{aligned}\bar{B}_{MN}(\omega) \simeq G_{MN}(\omega) \Big[ &\tilde{S}^{(3)}(\omega_2^N, \omega_1^M + \omega_2^N, \omega_2^N + \omega_{r2}\omega/\delta\omega_r) \\ &+ \tilde{S}^{(3)}(\omega_2^N, \omega_1^M + \omega_2^N, \omega_1^M - \omega_{r1}\omega/\delta\omega_r) \\ &+ \tilde{S}^{(3)}(\omega_2^N, -\omega_1^M + \omega_2^N + \omega_{r2}\omega/\delta\omega_r, \omega_2^N + \omega_{r2}\omega/\delta\omega_r) \\ &+ \tilde{S}^{(3)}(\omega_2^N, -\omega_1^M + \omega_2^N + \omega_{r2}\omega/\delta\omega_r, -\omega_1^M) \\ &+ \tilde{S}^{(3)}(\omega_2^N, -\omega_{r1}\omega/\delta\omega_r, \omega_1^M - \omega_{r1}\omega/\delta\omega_r) \\ &+ \tilde{S}^{(3)}(\omega_2^N, -\omega_{r1}\omega/\delta\omega_r, -\omega_1^M) \Big].\end{aligned} \tag{48}$$

Using this result, $\bar{\bar{B}}_{MN}(t)$ introduced in Eq. (33) can be calculated by inverse Fourier transform

$$\bar{\bar{B}}_{MN}(t) = \frac{1}{2\pi} \int_{-\infty}^{\infty} d\omega \bar{B}_{MN}(\omega) e^{-i\omega t}. \tag{49}$$

Because the frequency variable $\omega$ appears in the second and the third arguments of the response functions in Eq. (48), the inverse Fourier transform can be carried out analytically only for some of the terms. In particular, as pointed out in section 2.B, the sixth term is expected to make the largest contribution to $\bar{B}_{MN}(\omega)$ considering the time ordering of pulses. In addition, according to Eq. (45), the extremal values of $M$ and $N$ to consider are about $\pm 10^5$ and, therefore, $\omega_1^M$ and $\omega_2^N$ appearing in Eq. (48) are rather narrowly distributed around the mid-points of $\omega_{c1}$ and $\omega_{c2}$, respectively. Then, the sixth term $G_{MN}(\omega)\tilde{S}^{(3)}(\omega_2^N,-\omega_{r1}\omega/\delta\omega_r,-\omega_1^M)$ represents a rephasing quantum transition pathway consistent with the signal wavevector $\mathbf{k}_s = -\mathbf{k}_1 + \mathbf{k}_2 + \mathbf{k}_3$ because the signs of the two frequencies in the first and third arguments of $\tilde{S}^{(3)}$ are opposite. From these considerations, we hereafter focus on the sixth term and write its inverse Fourier transform as

$$\bar{\bar{B}}_{MN}^{[6]}(t) = \frac{1}{2\pi}\int_{-\infty}^{\infty}d\omega \bar{B}_{MN}^{[6]}(\omega)e^{-i\omega t} \\ \simeq \frac{1}{2\pi}\int_{-\infty}^{\infty}d\omega\, e^{-i\omega t}G_{MN}(\omega)\tilde{S}^{(3)}(\omega_2^N,-\omega_{r1}\omega/\delta\omega_r,-\omega_1^M)$$
(50)

where the superscript "[6]" denotes the contribution of the sixth term. We also introduce the inverse Fourier transform $g_{MN}(t)$ of $G_{MN}(\omega)$ defined as

$$g_{MN}(t) = \frac{1}{2\pi}\int_{-\infty}^{\infty}d\omega\, G_{MN}(\omega)e^{-i\omega t} \\ = \frac{\omega_{r1}^2\omega_{r2}^2}{16\pi^4}\sqrt{\frac{1}{4\pi a}}e^{\frac{\beta^2+4\alpha\gamma}{4\alpha}}e^{-i\beta t/(2\alpha)}e^{-t^2/(4\alpha)}$$
(51)

where $\alpha$, $\beta$, and $\gamma$ are given as

$$\alpha = \frac{\sigma_1^2\omega_{r1}^2 + \sigma_2^2\omega_{r2}^2}{2(\delta\omega_r)^2} \\ \beta = \frac{M\sigma_1^2\omega_{r1}^2 - N\sigma_2^2\omega_{r2}^2}{\delta\omega_r} \\ \gamma = -M^2\sigma_1^2\omega_{r1}^2 - N^2\sigma_2^2\omega_{r2}^2$$
(52)

Then, Eq. (50) can be written as follows

$$\bar{\bar{B}}_{MN}^{[6]}(t) = \frac{\delta\omega_r}{\omega_{r1}}\int_{-\infty}^{\infty}d\tau\, g_{MN}(t-\tau)\bar{S}(\omega_2^N,-(\delta\omega_r/\omega_{r1})\tau,-\omega_1^M) \\ = \frac{\delta\omega_r}{\omega_{r1}}\int_0^{\infty}d\tau\, g_{MN}(t+\tau)\bar{S}(\omega_2^N,(\delta\omega_r/\omega_{r1})\tau,-\omega_1^M)$$
(53)

where $\bar{S}(\omega_3,t_2,\omega_1)$ is the 2D Fourier transform of the time domain response function $S^{(3)}(t_3,t_2,t_1)$ with respect to $t_1$ and $t_3$. In Eq. (53), we have utilized the convolution theorem for two functions with scaled variables:

$$F(a\omega)G(c\omega) = \int_{-\infty}^{\infty}dt\, f(t)e^{ia\omega t}\int_{-\infty}^{\infty}dt'\, g(t')e^{ic\omega t'} \\ = \frac{1}{|ac|}\int_{-\infty}^{\infty}d\tau' e^{i\omega\tau'}\int_{-\infty}^{\infty}d\tau\, f(\tau/a)g\left(\frac{\tau'-\tau}{c}\right)$$
(54)

and the condition that $\bar{S}(\omega_3,t_2,\omega_1)=0$ for $t_2<0$ was also used [8]. We can see that, for the sixth term, the 2D signal can be expressed as the convolution of the function $\bar{S}(\omega_3,t_2,\omega_1)$ with a Gaussian function. This indicates that the waiting time dependence of the response function could be extracted from the $t$-dependence of the signal but its amplitude might be modulated by the Gaussian pulse envelope. We also note that this effective waiting time $T_w$ is scaled down from the laboratory detection time $t$ by the frequency down-conversion factor $\delta\omega_r/\omega_{r1}$ as follows

$$T_w = (\delta\omega_r/\omega_{r1})t. \tag{55}$$

The 2D spectrum due to the sixth term, $S_{2D}^{[6]}(t;\omega_{\tau 1},\omega_{\tau 2})$, can be obtained from the following amplitude function which is derived from Eqs. (36), (37) and (53)

$$\begin{aligned} F^{[6]}(t;\omega_{\tau 1},\omega_{\tau 2}) &= \frac{\pi^3}{\delta\omega_r}\overline{\overline{B}}_{\overline{M}(\omega_{\tau 1})\overline{N}(\omega_{\tau 2})}^{[6]}(-t) \\ &= \frac{\pi^3}{\omega_{r1}}\int_0^\infty d\tau\, g_{\overline{M}(\omega_{\tau 1})\overline{N}(\omega_{\tau 2})}(-t+\tau)\overline{S}(\omega_{c2}+\overline{N}(\omega_{\tau 2})\omega_{r2},(\delta\omega_r/\omega_{r1})\tau,-\omega_{c1}-\overline{M}(\omega_{\tau 1})\omega_{r1}) \\ &= \frac{\pi^3}{\omega_{r1}}\int_0^\infty d\tau\, g_{\overline{M}(\omega_{\tau 1})\overline{N}(\omega_{\tau 2})}(-t+\tau)\overline{S}(\omega_{\tau 2},(\delta\omega_r/\omega_{r1})\tau,\omega_{\tau 1}) \end{aligned} \tag{56}$$

by applying Eq. (38) as follows

$$\begin{aligned} &S_{2D}^{[6]}(t;\omega_{\tau 1},\omega_{\tau 2}) \\ &= \mathrm{Im}\left[F^{[6]}(t;\omega_{\tau 1},\omega_{\tau 2})+F^{[6]}(t;-\omega_{\tau 1},-\omega_{\tau 2})\right]-i\,\mathrm{Re}\left[F^{[6]}(t;\omega_{\tau 1},\omega_{\tau 2})-F^{[6]}(t;-\omega_{\tau 1},-\omega_{\tau 2})\right]. \end{aligned} \tag{57}$$

## 3. Model Calculation

### A. Model Description

We demonstrate the theory of the DFC 3PPE developed above by calculating the derived 2D spectrum using a model response function for a two-level system (2LS). In this model composed of the ground state $g$ and the excited state $e$ (See Sec. 5.2-5.4, 7.6 of Ref. [8]), the third-order response function has the following expression

$$S^{(3)}(t_3,t_2,t_1) = \left(\frac{i}{\hbar}\right)^3 \theta(t_3)\theta(t_2)\theta(t_1)\sum_{n=1}^{4}\left[R_n(t_3,t_2,t_1)-R_n^*(t_3,t_2,t_1)\right] \tag{58}$$

in terms of $R_n$ given by

$$\begin{aligned} R_1(t_3,t_2,t_1) &= \mu^4 e^{-i\omega_{eg}t_3-i\omega_{eg}t_1}F_1(t_3,t_2,t_1), \\ R_2(t_3,t_2,t_1) &= \mu^4 e^{-i\omega_{eg}t_3+i\omega_{eg}t_1}F_2(t_3,t_2,t_1), \\ R_3(t_3,t_2,t_1) &= \mu^4 e^{-i\omega_{eg}t_3+i\omega_{eg}t_1}F_3(t_3,t_2,t_1), \\ R_4(t_3,t_2,t_1) &= \mu^4 e^{-i\omega_{eg}t_3-i\omega_{eg}t_1}F_4(t_3,t_2,t_1) \end{aligned} \tag{59}$$

Here, $\mu = |\mu_{eg}|$ is the absolute value of the transition dipole moment, $\omega_{eg}$ is the ensemble-averaged transition frequency, and $F_n(t_3,t_2,t_1)$ is the line shape function given as follows under the second-order cumulant expansion approximation [26] and the short-time approximation for the coherence time variables $t_1$ and $t_3$ [29]

$$\ln F_n(t_3,t_2,t_1) = f_n(t_2) - \delta_n^2(t_2)t_1^2/2 - \Delta_n^2(t_2)t_3^2/2 + H_n(t_2)t_1t_3 + iQ_n(t_2)t_3. \tag{60}$$

Explicit expressions of $f_n(t)$, $\delta_n(t)$, $\Delta_n(t)$, $H_n(t)$, and $Q_n(t)$ can be found in Eqs. (5.36)-(5.41) of Ref. [8] for a general multi-level system. For the current 2LS, they are simplified as follows

$$
\begin{aligned}
&f_n(t) = 0 \quad (n=1,\cdots,4), \\
&\delta_n^2(t) = \Delta_n^2(t) = C_{ee}(0) \quad (n=1,\cdots,4), \\
&H_1(t) = -H_2(t) = -H_3(t) = H_4(t) = -\text{Re}[C_{ee}(t)], \\
&Q_1(t) = Q_2(t) = \text{Im}\left[2\bar{C}_{ee}(0) - \bar{C}_{ee}(t) - \bar{C}_{ee}(-t)\right], \\
&Q_3(t) = Q_4(t) = 0
\end{aligned}
\tag{61}
$$

where the frequency-frequency time correlation functions (FFCFs) and related quantities are defined as follows

$$
\begin{aligned}
C_{ab}(t) &= \langle \delta V_{ag}(t)\delta V_{bg}(0)\rangle_{\text{Bath}}/\hbar^2 = \langle \delta\omega_{ag}(t)\delta\omega_{bg}(0)\rangle_{\text{Bath}}, \\
g_{ab}(t) &= \int_0^t d\tau_1 \int_0^{\tau_1} d\tau_2 C_{ab}(\tau_2) = \int_0^t d\tau_1 (t-\tau_1) C_{ab}(\tau_1), \\
\bar{C}_{ab}(t) &= \int_0^t d\tau C_{ab}(\tau) = \frac{dg_{ab}(t)}{dt}
\end{aligned}
\tag{62}
$$

in terms of the fluctuations in the energy gap, $\delta V_{ag}(t) = V_{ag}(t) - \langle V_{ag}(t)\rangle_{\text{Bath}}$.

To specify the FFCF, we introduce the spectral density $\rho(\omega)$ representing the chromophore-bath couplings [24] as follows

$$
\begin{aligned}
\rho(\omega) &= \frac{\tilde{C}''(\omega)}{\pi\omega^2}, \\
\tilde{C}''(\omega) &= i\int_{-\infty}^{\infty} dt\, C_I(t) e^{i\omega t} = i\tilde{C}_I(\omega) = -\text{Im}\left[\tilde{C}_I(\omega)\right]
\end{aligned}
\tag{63}
$$

where $C_I(t)$ is the imaginary part of $C(t) \equiv C_{ee}(t)$. Note that $\tilde{C}''(\omega)$ and $\rho(\omega)$ are real and odd functions of $\omega$ due to the general relation $C(-t) = C^*(t)$ for a quantum time correlation function $C(t)$. The FFCF can be written in terms of $\rho(\omega)$ as follows [8]

$$
C(t) = \int_0^\infty d\omega\, \omega^2 \coth(\beta\hbar\omega/2)\cos(\omega t)\rho(\omega) - i\int_0^\infty d\omega\, \omega^2 \sin(\omega t)\rho(\omega) \tag{64}
$$

which stems from the detailed balance relation $\tilde{C}(-\omega) = e^{-\beta\hbar\omega}\tilde{C}(\omega)$ for the Fourier transform $\tilde{C}(\omega)$ of $C(t)$ [26] where $\beta^{-1} = k_B T$ with the Boltzmann constant $k_B$ and the absolute temperature $T$. The quantities in Eq. (61) can then be expressed in terms of $\rho(\omega)$ as follows

$$
\begin{aligned}
\Omega^2 &\equiv C(0) = \delta_n^2(t) = \Delta_n^2(t) \\
&= \int_0^\infty d\omega\, \omega^2 \coth(\beta\hbar\omega/2)\rho(\omega) \quad (n=1,\cdots,4), \\
H(t) &\equiv \text{Re}[C(t)] = -H_1(t) = H_2(t) = H_3(t) = -H_4(t) \\
&= \int_0^\infty d\omega\, \omega^2 \coth(\beta\hbar\omega/2)\cos(\omega t)\rho(\omega), \\
Q(t) &\equiv -Q_1(t)/2 = -Q_2(t)/2 \\
&= \int_0^\infty d\omega\, \omega \cos(\omega t)\rho(\omega) - \lambda/\hbar
\end{aligned}
\tag{65}
$$

where the solvent reorganization energy $\lambda$ is defined as

$$\lambda = \hbar \int_0^\infty d\omega\, \omega \rho(\omega). \tag{66}$$

Now, the four response function components in Eq. (59) can be written as

$$\begin{aligned}
R_1(t_3, t_2, t_1) &= \mu^4 e^{-i\omega_{eg} t_1 - i[\omega_{eg} + 2Q(t_2)]t_3} e^{-\Omega^2 t_1^2/2 - \Omega^2 t_3^2/2 - H(t_2) t_1 t_3}, \\
R_2(t_3, t_2, t_1) &= \mu^4 e^{i\omega_{eg} t_1 - i[\omega_{eg} + 2Q(t_2)]t_3} e^{-\Omega^2 t_1^2/2 - \Omega^2 t_3^2/2 + H(t_2) t_1 t_3}, \\
R_3(t_3, t_2, t_1) &= \mu^4 e^{i\omega_{eg} t_1 - i\omega_{eg} t_3} e^{-\Omega^2 t_1^2/2 - \Omega^2 t_3^2/2 + H(t_2) t_1 t_3}, \\
R_4(t_3, t_2, t_1) &= \mu^4 e^{-i\omega_{eg} t_1 - i\omega_{eg} t_3} e^{-\Omega^2 t_1^2/2 - \Omega^2 t_3^2/2 - H(t_2) t_1 t_3}
\end{aligned} \tag{67}$$

using the quantities in Eq. (65).

In the model calculation, we employ the Ohmic spectral density with an exponential cutoff function [8] as follows

$$\rho(\omega) = \frac{\lambda}{\hbar \omega_0} \frac{e^{-\omega/\omega_0}}{\omega} \quad (\omega \geq 0) \tag{68}$$

which yields

$$\begin{aligned}
\Omega^2 &\simeq \frac{2\lambda}{\beta \hbar^2}, \\
H(t) &\simeq \frac{2\lambda}{\beta \hbar^2} \frac{1}{1 + \omega_0^2 t^2}, \\
Q(t) &= -\frac{\lambda}{\hbar} \frac{\omega_0^2 t^2}{1 + \omega_0^2 t^2}
\end{aligned} \tag{69}$$

where we have taken the high temperature limit ($\beta \to 0$) for $\Omega^2$ and $H(t)$ which is expected to introduce at most 1 % error at room temperature with the choice of $\omega_0 = 30\,\mathrm{cm}^{-1}$.

## B. Model Response Function

To evaluate the 2D spectrum $S_{2D}^{[6]}(t; \omega_{\tau 1}, \omega_{\tau 2})$ in Eq. (57), we first consider the 2D Fourier transform $\bar{S}(\omega_3, t_2, \omega_1)$ of the third order response function

$$\bar{S}(\omega_3, t_2, \omega_1) = \int_0^\infty dt_3 \int_0^\infty dt_1\, S^{(3)}(t_3, t_2, t_1) e^{i(\omega_3 t_3 + \omega_1 t_1)}. \tag{70}$$

Since $S^{(3)}(t_3, t_2, t_1)$ is given by Eqs. (58) and (67) in the present model, $\bar{S}(\omega_3, t_2, \omega_1)$ is composed of integrals of the following form

$$\begin{aligned}
I(\chi_3, \chi_1) &= \int_0^\infty dt_3 \int_0^\infty dt_1\, e^{i\chi_1 t_1 + i\chi_3 t_3} e^{-\Omega^2 t_1^2/2 - \Omega^2 t_3^2/2 - H t_1 t_3} \\
&= 2 \int_0^\infty dz_1\, e^{i(\chi_1 + \chi_3) z_1/\sqrt{2}} e^{-\lambda_1 z_1^2/2} \int_0^{z_1} dz_3 \cos\left[(\chi_1 - \chi_3) z_3/\sqrt{2}\right] e^{-\lambda_3 z_3^2/2}
\end{aligned} \tag{71}$$

where we have transformed the integration variables in the second step to eliminate the cross term in the exponent as shown in Appendix. Taking the upper limit of the inner integral to infinity, we can obtain an approximate expression of this integral as follows

$$I(\chi_3, \chi_1) \cong 2\int_0^\infty dz_1 e^{i(\chi_1+\chi_3)z_1/\sqrt{2}} e^{-\lambda_1 z_1^2/2} \int_0^\infty dz_3 \cos\left[(\chi_1-\chi_3)z_3/\sqrt{2}\right] e^{-\lambda_3 z_3^2/2}$$
$$= \frac{\pi}{\sqrt{\lambda_1 \lambda_3}} e^{-(\chi_1-\chi_3)^2/(4\lambda_3)} \left[ e^{-(\chi_1+\chi_3)^2/(4\lambda_1)} + \frac{2i}{\sqrt{\pi}} D\left(\frac{\chi_1+\chi_3}{2\sqrt{\lambda_1}}\right) \right] \quad (72)$$

where $D(x) = (1/2)\int_0^\infty d\tau e^{-\tau^2/4} \sin x\tau = e^{-x^2} \int_0^x d\tau e^{\tau^2}$ is the Dawson function [30]. Compared to some previous approaches [8], the response function evaluated with Eq. (72) would be more desirable because it can capture the correlation between the two frequency variables.

The eight terms in Eq. (58) can be expressed in the form of Eq. (72) with different $\chi_{1,3}$ and $\lambda_{1,3}$ as follows

$$\begin{aligned}
R_1: &\quad \chi_1 = \omega_1 - \omega_{eg}, \quad \chi_3 = \omega_3 - \omega_{eg} - 2Q(t_2), \\
R_2: &\quad \chi_1 = \omega_1 + \omega_{eg}, \quad \chi_3 = \omega_3 - \omega_{eg} - 2Q(t_2), \\
R_3: &\quad \chi_1 = \omega_1 + \omega_{eg}, \quad \chi_3 = \omega_3 - \omega_{eg}, \\
R_4: &\quad \chi_1 = \omega_1 - \omega_{eg}, \quad \chi_3 = \omega_3 - \omega_{eg}, \\
R_1^*: &\quad \chi_1 = \omega_1 + \omega_{eg}, \quad \chi_3 = \omega_3 + \omega_{eg} + 2Q(t_2), \\
R_2^*: &\quad \chi_1 = \omega_1 - \omega_{eg}, \quad \chi_3 = \omega_3 + \omega_{eg} + 2Q(t_2), \\
R_3^*: &\quad \chi_1 = \omega_1 - \omega_{eg}, \quad \chi_3 = \omega_3 + \omega_{eg}, \\
R_4^*: &\quad \chi_1 = \omega_1 + \omega_{eg}, \quad \chi_3 = \omega_3 + \omega_{eg}.
\end{aligned} \quad (73)$$

$$\begin{aligned}
R_1, R_4, R_1^*, R_4^*: &\quad \lambda_1 = \Omega^2\left[2 + (\hbar/\lambda)Q(t_2)\right], \quad \lambda_3 = -\Omega^2(\hbar/\lambda)Q(t_2), \\
R_2, R_3, R_2^*, R_3^*: &\quad \lambda_1 = -\Omega^2(\hbar/\lambda)Q(t_2), \quad \lambda_3 = \Omega^2\left[2 + (\hbar/\lambda)Q(t_2)\right].
\end{aligned} \quad (74)$$

From Eq. (69), we note the following inequalities
$$\begin{aligned}
-\lambda/\hbar &\leq Q(t_2) \leq 0, \\
0 &\leq H(t_2) \leq \Omega^2, \\
\Omega^2 &\leq \Omega^2\left[2 + (\hbar/\lambda)Q(t_2)\right] \leq 2\Omega^2, \\
0 &\leq -\Omega^2(\hbar/\lambda)Q(t_2) \leq \Omega^2.
\end{aligned} \quad (75)$$

which show that both $\lambda_1$ and $\lambda_3$ are non-negative. With the following choice of parameters

$$\begin{aligned}
\omega_{eg} &= 10000 \text{ cm}^{-1} = 299.8 \text{ THz}, \\
\lambda/\hbar &= 500 \text{ cm}^{-1}, \\
\omega_0 &= 30 \text{ cm}^{-1}, \\
k_B T/\hbar &= 1/(\beta\hbar) = 208.5 \text{ cm}^{-1} \; (T = 300 \text{ K}),
\end{aligned} \quad (76)$$

we obtain

$$\begin{aligned}
\Omega &= 456.6 \text{ cm}^{-1}, \\
-500 \text{ cm}^{-1} &\leq Q(t_2) \leq 0 \text{ cm}^{-1}.
\end{aligned} \quad (77)$$

From Eqs. (72), (75) and (77), we can see that $\bar{S}(\omega_3, t_2, \omega_1)$ is the sum of eight terms that have peaks at or around one of $(\pm\omega_{eg}, \pm\omega_{eg})$ with width on the order of 500 cm$^{-1}$. For example, since $\chi_1 \pm \chi_3$ should be small to produce large signal according to Eq. (72), the signal from $R_1$ would appear in the first quadrant near $(\omega_{eg}, \omega_{eg} + 2Q(t_2))$ according to Eq. (73). Proceeding analogously, we can establish the location of signals from the eight terms of $\bar{S}(\omega_3, t_2, \omega_1)$ as follows: $R_1$ and $R_4$ in the first quadrant, $R_2$ and $R_3$ in the second, $R_1^*$ and $R_4^*$ in the third, and $R_2^*$ and $R_3^*$ in the fourth. On the other hand, terms of $\bar{S}(-\omega_3, t_2, -\omega_1)$ would appear as follows: $R_1$ and $R_4$ in the third quadrant, $R_2$ and $R_3$ in the fourth, $R_1^*$ and $R_4^*$ in the second, and $R_2^*$ and $R_3^*$ in the first. Since the signal amplitude is modulated by $g_{\bar{M}(\omega_{r1})\bar{N}(\omega_{r2})}(t)$ according to Eq. (56), we also need to consider its behavior. From Eqs. (37), (51), and (52), we can see that $g_{\bar{M}(\omega_{r1})\bar{N}(\omega_{r2})}(t)$ is non-negligible only in the second quadrant and $g_{\bar{M}(-\omega_{r1})\bar{N}(-\omega_{r2})}(t)$, in the fourth quadrant. Therefore, the 2D spectrum in Eq. (57) would arise only from the rephasing pathways $R_2$ and $R_3$ and appear in the second and fourth quadrants with inversion symmetry with respect to the origin.

*C. Model 2D Spectrum*

Based on the model described above, we calculate the 2D spectrum expected from the DFC 3PPE experiment. We first note that the summation over the comb indices $M$ and $N$ appearing in Eq. (26) is not necessary because of the two Dirac delta functions obtained in Eq. (30) that impose the relation between these indices and the frequency variables $(\omega_{r1}, \omega_{r2})$ as given by Eq. (37). In addition, the summation over the index $L$ in Eq. (26) is replaced by the integral over $\omega$ in Eq. (33). Therefore, we will directly evaluate the 2D spectrum using Eq. (57) without intermediate calculation of time-domain interferogram.

In the computation, we employ the molecular parameters defined in Eqs. (76) and (77) and the following comb field parameters

$$\sigma_1 = \sigma_2 = 10 \text{ fs (FWHM bandwidth = 23 THz or 1270 cm}^{-1}),$$
$$\bar{\omega}_c = (\omega_{c1} + \omega_{c2})/2 = 10000 \text{ cm}^{-1} = 299.8 \text{ THz},$$
$$\delta\omega_c = \omega_{c1} - \omega_{c2} = 1.5 \times 10^{-3} \text{ cm}^{-1} = 44.97 \text{ MHz}, \qquad (78)$$
$$\omega_{r1} = 0.01 \text{ cm}^{-1} = 299.8 \text{ MHz},$$
$$\delta\omega_r = 1.5 \times 10^{-9} \text{ cm}^{-1} = 44.97 \text{ Hz}.$$

The computation was performed using the GNU Octave program [31] which provides means to evaluate the Dawson function and the convolution integral in Eq. (56). The 2D spectra obtained with these parameters are displayed in Fig. 1 for the effective waiting times $T_w = 100, 200, 300, 400, 500$ fs, corresponding to laboratory detection time $t = 0.6667, 1.333, 2.000, 2.666, 3.333$ ns, respectively. Since the spectrum appears in the second and fourth quadrants with inversion symmetry as shown above, we only plot the unique spectrum in the second quadrant as a function of $(-\omega_{r1}, \omega_{r2})$. The real parts of the spectra in Figure 1 (a)-(e) show characteristic absorptive spectral lineshape, and the imaginary parts in Figure 1 (f)-(j) exhibit dispersive lineshape with nodal line between the positive and negative signals. As the waiting time $T_w$ increases, the signal loses its diagonal correlation and also the amplitude decreases due to dephasing. The peak positions are slightly red-shifted along the $\omega_{r2}$ axis, which arises from the solvation dynamics described by $Q(t_2)$. These numerical calculation results show that the DFC 3PPE spectroscopy reproduces all the spectral features expected in a conventional 2D photon echo spectrum obtained by using a single mode-locked femtosecond laser.

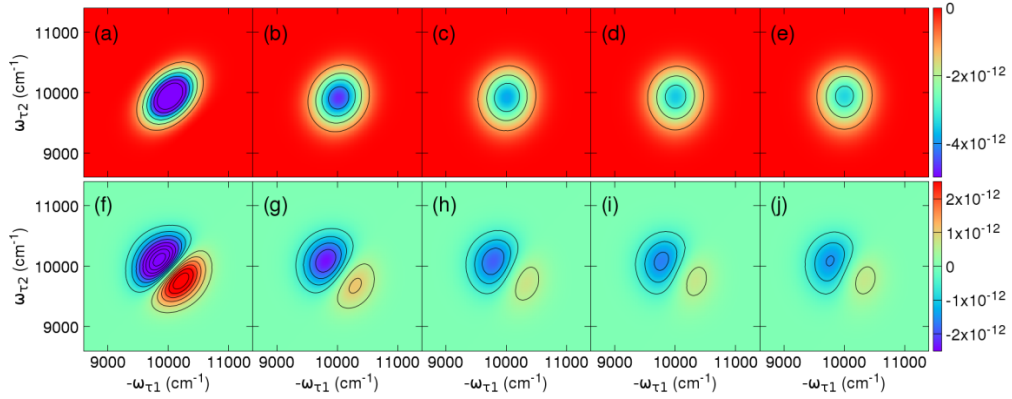

Fig. 1. Theoretical 2D spectra of DFC 3PPE of a two-level system. (a)-(e) Real parts at population time $T_w$ = 100, 200, 300, 400, 500 fs, respectively. (f)-(j) Imaginary parts at $T_w$ = 100, 200, 300, 400, 500 fs, respectively. The contour lines are drawn in increments of $1\times 10^{-12}$ in (a)-(e), and in increments of $5\times 10^{-13}$ in (f)-(j).

## 4. Summary and a Few Concluding Remarks

In summary, we have presented a theory of DFC 3PPE spectroscopy to understand the expected spectral features in terms of the underlying third order response function. The theory is an extension of our previous work on the photon echo spectroscopy that employs two optical frequency comb lasers. Here, we have shown that the detection time of the heterodyne-detected signal interferogram is transparently related to the population or waiting time of 3PPE spectroscopy by the frequency down-conversion factor of DFC. In addition, the calculated 2D model spectra closely resembles the 2D spectra expected from conventional 3PPE measurements employing a single mode-locked laser and a non-collinear spectral interferometric detection scheme. It is therefore shown that the DFC 3PPE spectroscopy could be a robust alternative to conventional 3PPE methods.

In the experimental configuration considered here, the ASOPS enabled by the repetition frequency offset $\delta\omega_r$ appears in the measurement time of the interferogram which is related to the population time. It would be possible to devise modified configurations which could better exploit the ASOPS feature of DFC. For instance, by temporally interlocking the pulse trains in an alternating order, i.e., comb1-comb2-comb1-comb2 or comb1-comb2-comb2-comb1, the ASOPS could be realized for the coherence times between the first and second pulses or between the third pulse and the detection time. It would also be worthwhile to explore the applicability of DFC to 2D spectroscopy in the collinear pump-probe geometry [32, 33] that is another method of widespread use. The theory presented in this paper could be applied to these cases and promote further development of DFC nonlinear spectroscopy.

## Appendix. An Integral Identity For Response Function Evaluation

Here, we derive an integral identity for $I(\chi_3, \chi_1)$ in Eq. (71)

$$I(\chi_3, \chi_1) = \int_0^\infty dt_3 \int_0^\infty dt_1 e^{i\chi_1 t_1 + i\chi_3 t_3} e^{-\Omega^2 t_1^2/2 - \Omega^2 t_3^2/2 - H t_1 t_3} \qquad (79)$$

which is a 2D half Fourier transform of a Gaussian with bilinear coupling. If we introduce a symmetric real matrix **A** and a column vector **t** as

$$\mathbf{A} = \begin{pmatrix} \Omega^2 & H \\ H & \Omega^2 \end{pmatrix}, \tag{80}$$
$$\mathbf{t}^T = (t_1, t_3),$$

$I(\chi_3, \chi_1)$ can be written as

$$I(\chi_3, \chi_1) = \int_0^\infty dt_3 \int_0^\infty dt_1 e^{i\chi_1 t_1 + i\chi_3 t_3} e^{-\mathbf{t}^T \mathbf{A} \mathbf{t}/2}. \tag{81}$$

There exists an orthogonal matrix **M** that diagonalizes **A** as follows

$$\mathbf{D} = \mathrm{diag}(\lambda_1, \lambda_2) = \mathbf{M}^T \mathbf{A} \mathbf{M}, \tag{82}$$
$$\mathbf{z} = \mathbf{M}^T \mathbf{t}.$$

These quantities can be found from Eq. (80) as

$$\mathbf{M} = \frac{1}{\sqrt{2}} \begin{pmatrix} 1 & 1 \\ 1 & -1 \end{pmatrix},$$
$$\mathbf{D} = \begin{pmatrix} \lambda_1 & 0 \\ 0 & \lambda_3 \end{pmatrix} = \begin{pmatrix} \Omega^2 + H & 0 \\ 0 & \Omega^2 - H \end{pmatrix}, \tag{83}$$
$$\mathbf{z} = \begin{pmatrix} z_1 \\ z_3 \end{pmatrix} = \frac{1}{\sqrt{2}} \begin{pmatrix} t_1 + t_3 \\ t_1 - t_3 \end{pmatrix}.$$

Then, $I(\chi_3, \chi_1)$ becomes

$$\begin{aligned} I(\chi_3, \chi_1) &= \int_0^\infty dt_3 \int_0^\infty dt_1 e^{i\chi_1 t_1 + i\chi_3 t_3} e^{-\mathbf{z}^T \mathbf{D} \mathbf{z}/2} \\ &= \int_0^\infty dz_1 \int_{-z_1}^{z_1} dz_3 e^{i[(\chi_1 + \chi_3)z_1 + (\chi_1 - \chi_3)z_3]/\sqrt{2}} e^{-\lambda_1 z_1^2/2 - \lambda_3 z_3^2/2} \\ &= \int_0^\infty dz_1 e^{i(\chi_1 + \chi_3)z_1/\sqrt{2}} e^{-\lambda_1 z_1^2/2} \int_{-z_1}^{z_1} dz_3 e^{i(\chi_1 - \chi_3)z_3/\sqrt{2}} e^{-\lambda_3 z_3^2/2} \\ &= 2\int_0^\infty dz_1 e^{i(\chi_1 + \chi_3)z_1/\sqrt{2}} e^{-\lambda_1 z_1^2/2} \int_0^{z_1} dz_3 \cos\left[(\chi_1 - \chi_3)z_3/\sqrt{2}\right] e^{-\lambda_3 z_3^2/2} \end{aligned} \tag{84}$$

where the change of integration limits in the second step can be understood from Figure A.

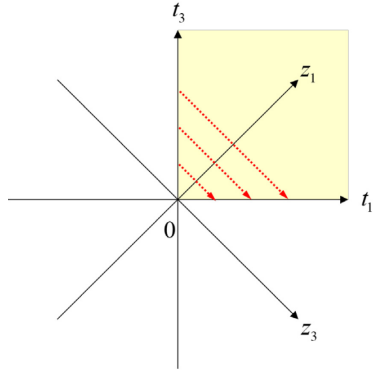

Fig. A. Integration ranges of $I(\chi_3, \chi_1)$ in the transformed coordinate $(z_1, z_3) = (t_1 + t_3, t_1 - t_3)/\sqrt{2}$ indicated by dotted arrows.

## Funding

We thank the financial support by Institute for Basic Science (IBS) of Korea (IBS-R023-D1).

## Disclosure

The authors declare that there are no conflicts of interest related to this article.